\documentclass[floatfix,showpacs,amsmath,amssymb,letterpaper,groupaddresses,superscriptaddress]{article}
\usepackage{latexsym}
\usepackage{epsfig}

\usepackage{amsmath}
\usepackage{amssymb}
\usepackage{graphicx}
\usepackage{times}
\usepackage[section]{placeins}
\usepackage{caption}
\usepackage{subcaption}

\usepackage{color}

\def\a{\alpha}
\def\b{\beta}

\def\be{\begin{equation}}
\def\ee{\end{equation}}
\def\ba{\begin{eqnarray}}
\def\ea{\end{eqnarray}}
\def\la{\langle}
\def\ra{\rangle}
\def\a{\alpha}

\def\b{\beta}

\def\h{\hskip 1cm}
\def\hh{\hskip 2cm}

\begin{document}

\vspace{4cm}
\begin{center}{\Large \bf Secure alignment of coordinate systems by using quantum correlation}\\
\vspace{2cm}
\vspace{2cm}

F . Rezazadeh,$^1$\h A . Mani,$^2$\h V . Karimipour.$^1$\\
\vspace{1cm} $^1$ Department of Physics, Sharif University of Technology, P.O. Box 11155-9161, Tehran, Iran.\\
$^2$ Department of Sience Engineering, University of Tehran, Tehran, Iran.\\

\end{center}

\begin{abstract}
We show that two parties far apart can use shared entangled states and classical communication to align their  coordinate systems with a very high fidelity. Moreover, compared with  previous methods proposed for such a task, i.e. sending parallel or anti-parallel pairs or groups of spin states, our method has the extra advantages of using single qubit measurements and also being secure, so that third parties do not extract any information about the aligned coordinate system established between the two parties. The latter property is important in many other quantum information protocols in which measurements inevitably play a significant role.

\end{abstract}

\vskip 2cm

\section{Introduction}
Almost any  protocol in quantum communication between two or more parties, requires measurements in bases which are agreed upon by the parties involved \cite{quantum teleportation, dense coding, key distribution1, key distribution2}. This in turn requires that they establish aligned reference frames between them with arbitrary precision.  It has been realized that spatial direction is a type of information named 'unspeakable quantum information' \cite{unspeakable information} which can not be transmitted by sending classical bits unless the sender (Alice) and the receiver (Bob) have a common coordinate system. Instead, physical objects, e.g. photons must be sent \cite{ Rudolph} to convey this information.\\

The problem of setting a reference frame between two apart parties can be reduced to the problem of sharing three mutually orthogonal directions.
Some methods of direction sharing are based on transmission of spin states (qubits), followed by single or multi-qubit measurements which are performed by the receiver \cite{Gisin and Popescu,Massar and Popescu,Began,Peres and Scudo}.
For example Gisin and Popescu investigate the case when many pairs of parallel or anti-parallel spins are transmitted from Alice to Bob and show that a higher fidelity is achieved when the two spins are anti-parallel to each other \cite{Gisin and Popescu}. In \cite{Gisin and Popescu}, the strategy of Bob for guessing the direction of spins sent to him, is based on using a specific measurement of two-qubit entangled states. The in-equivalence with the case of two parallel spins is justified by noting that there is no universal NOT machine which can turn any unknown pair of anti-parallel spins into a pair of parallel spins.  It is shown in \cite{Massar and Popescu} that in such a case, the average optimal fidelity of direction sharing is equal to $ \dfrac{N+1}{N+2} $ . This line of thought has been further persued in \cite{S.n}, where it has been shown that for conveying the direction ${\bf n}$, Alice can encode it into a specific eigenstate of the operator ${\bf S}\cdot {\bf n}$ where ${\bf S}$ is the total spin operator and send it to Bob who will discern ${\bf n}$ with a collective measurement.\\

In another interesting approach,  Alice and Bob find the unitary operation that rotates the (rigid) frame of Bob to align it with the (rigid) frame of Alice \cite{Referee2001, Referee2004}. Thus in one go, the two frames are aligned. The method is based on sharing a $2N$-qubit highly entangled state upon which collective measurements are done by Bob after Alice has sent her $N$-qubit share to him. Clearly this method while being optimal in a theoretical sense, is  experimentally demanding . \\

In all these works and many other similar works \cite{optimal measurment1,optimal measurment2,optimal measurment3,Massar}  the question of secrecy of the directions has not been considered and for that reason the question of a possible role that  shared entangled states can play for such a goal has not been discussed. In view of the role that any quantum communication protocol is based on measurements in aligned coordinate systems, it is natural to demand that such alignment be made in a  secure and secret way so that only the legitimate parties know the directions and nobody else. To the best of our knowledge, there is only few works about the problem of security of reference frames , a notable example being \cite{SecretReferenceFrame}, where  Alice and Bob share a classical string of bits to achieve security. \\

In this paper we introduce a  method for direction sharing which uses only bi-partite entangled states and single -qubit measurements instead of multi-partite entangled states and collective measurements. Moreover no qubits are sent from one party to the other and there is no need for sharing strings of classical bits ascertain security. %in \cite{•} the authers use secret bits and quantum states to establish a secure reference frame beween Alice and Bob and they showed that if the quantum states (which Alice sends to Bob) are entangled the protocol achieves a better accuracy.
 In our protocol $N$ singlet states are shared between Alice and Bob and they do single qubit measurements in their specific but private directions. Finally they publicly announce the results of their measurements. From the correlations in these public data they can discern information about the relative angle between their directions of measurement and eventually align their coordinate systems in a precise way. Besides the secrecy in the common coordinate systems, our protocol has the advantage of using only single-qubit measurements, compared to multi-qubit measurements proposed in other methods  
 \cite{hydrogen atom1,hydrogen atom2,laboratory restriction1,laboratory restriction2,laboratory restriction3}.  \\
% There is also another work, \cite{Bahder}, where similar ideas had been put forward.  There, the author considers singlet states and uses mutual information between Alice's and Bob's measurment results for estimating the direction. However the method of \cite{Bahder} is based on a continuous family of  Bob trials in arbitrary directions until he finds the maximum mutual information. Therefore in that method a detailed comparison between different methods in terms of the consumed resources is not possible.\\
% 

We should also mention  \cite{Bahder}, where similar ideas to that of the present paper are  suggested and mutual information is used to align the two directions. There are however important differences between \cite{Bahder} and the present work. 
When the number of singlets are infinite, Bahder essentially tries to do an extensive (essentially continuous) search in order to find the direction which maximizes the mutual information. This certainly gives the aligned direction but is clearly infeasible for experiment. In the same limit, we do the alignment by determining the angles of one axis of Alice by measuring the correlations with those of Bob who measures his qubits in three different directions. So we do not need an exhaustive search over the sphere. When the number of pairs is finite, Bahder only goes as far as to estimate the angle between two directions using a Baeysian approach similar to that of us. However knowing only this angle, it is not evident around which axis the vectors should be rotated to be aligned.  Moreover the important point is that finite-$N$ fluctuations of correlations need to be carefully taken into account to extract useful geometrical information from these correlations. This problem will be dealt with in this paper. \\

In a sense our work is the converse of what is done in Quantum Key Distribution (QKD) \cite{key distribution1, key distribution2}, where Alice and Bob publicly announce their measurement bases but keep for themselves the results of measurements. Here they publicly announce their measurement results (the sequence of 0's and 1's or +'s and -'s) and from these public results they align their axes.  In the same way as QKD is secure, this protocol is also secure in the sense that Eavedroppers cannot gain information about the aligned directions.   \\

A special case of our method is when Alice and Bob already agree on a fixed direction, say the $z$ direction and they only want to align their $x$ and $y$ axes perpendicular to this axis. In this case, which we call the two dimensional case,  the protocol is simpler and can be done in just one step by estimating the angle between two directions and a complete alignment is achieved with a very high fidelity. In the general case where there is no a priori agreed direction or plane, two or three steps are needed, and again our method will lead to a very good estimate of the relative directions and hence alignment of coordinate systems. In both two and three dimensional cases, we first consider the ideal case where an infinite number of singlet states has been shared between Alice and Bob and then consider the realistic case where a finite nunmber of $N$ states has been shared in which case we obtain the fidelity of the protocol as a function of the number of shared singlets $N$. We will see that with few shared singlets, very high fidelities can be obtained. \\

The paper is organized as follows: in section (\ref{sec2}) we show how correlations of measurements of singlet states by two parties can lead to an estimation of the angle between directions of measurements. This is explained in two subsections, first for the ideal case where the number of singlets $N$ is infinite and then for the finite $N$ case, where we use a Baeysian approach to calculate the probabilities \cite{Bayesian}. In section (\ref{sec3}), we consider the geometrical problem of estimating a vector or direction by such measurements, and we compare our fidelities with that of others. The paper is concluded with a conclusion containing a discussion about the security of the protocol. \\

\section{Using entangled states to estimate the angle between two directions} \label{sec2}
Consider two parties, Alice and Bob, far apart from each other and share a number of singlet states
\begin{equation}
|\psi\ra=\dfrac{1}{\sqrt{2}}(|01\ra-|10\ra) ,
\end{equation}
where $|0\rangle $ and $|1\rangle $ are the eigenvectors of $ \vec\sigma.\vec z $, and $  \vec{\sigma} $ is the vector of pauli matrices:\\
\begin{equation}
\sigma_{x}=\left(
        \begin{array}{cc}
          0 & 1 \\
          1 &0 \\
        \end{array}
      \right), \hh \sigma_{y}=\left(
             \begin{array}{cc}
               0 &-i \\
               i & 0 \\
             \end{array}
           \right), \hh \sigma_{z}=\left(
             \begin{array}{cc}
               1 & 0 \\
               0 & -1 \\
             \end{array}
           \right).
           \end{equation}
           
To set up a shared direction in space, Alice and Bob use the correlations in their measurements to find how they should correct or rotate their axes of measurement. 
Alice and Bob measure their spins in two arbitrary directions only known to each of them separately. For example Bob measures his spin in his supposedly ${\bf z}$ direction and Alice measures her spins in a  direction which in the coordinate system of Bob is denoted by  ${\bf m}$ having an angle $\theta$ with ${\bf z}$. The aim of the experiment is to make the best estimate for this angle from the measurement of correlations. To this end, one of the parties, say Alice, publicly announces her  results in the form of a  sequence $(a_1, a_2, \cdots a_k, \cdots)$, where $a_i=\pm 1$. 
Bob compares this sequence with his own results $((b_1, b_{2}, \cdots b_k, \cdots)$, $b_i=\pm 1)$ and calculates the correlations between these two sequences, which for $N$ shared singlet pairs,  is given by:\\
\begin{equation}\label{Correlation a b def}
q_N=\frac{1}{N}\sum_{i=1}^{N}a_{i}\overline{b_{i}} ,
\end{equation}
This correlation function can be rewritten as:\\
\begin{equation}\label{Correlation Function def}
q_N=\frac{N_{+-}+N_{-+}-N_{++}-N_{--}}{N}=\frac{N_d-N_s}{N}=\frac{2N_d-N}{N} ,
\end{equation}
where $ N_{ab} $ denotes the number of the times that Alice obtains a value of $a$ and Bob obtains a value of $b$, and $N_{d} $ and $N_s$ are the number of times that Alice and Bob obtain different and the same results respectively.  It is evident that $ -1 \leq Q \leq 1 $ . If ${\bf m}$ and ${\bf z}$ are either parallel or anti-parallel, then the results will be fully correlated or fully anti-correlated and $\mid Q\mid=1$. For perpendicular directions, Bob obtains a value very close to $0$. From this correlation Bob can eventually determine the axis ${\bf m}$ of Alice in a three-step process. We will complete the idea in this section by first considering  the ideal case in which an infinite number of singlet pairs are shared amongst Alice and Bob and then considering the case of finite number of shared singlet pairs. 

\subsection{The case when an infinite number of pairs are shared} \label{infinite case}
In this case we will have:
\begin{equation}\label{infinitly correlation function}
Q_{\infty} = P_{+-}+P_{-+}-P_{++}-P_{--} ,
\end{equation}
where $ P_{ab} $ now denotes the probability of Alice obtaining a value of $a$ and Bob obtaining a value of $b$.  These probabilities are equal to:
\begin{equation}\label{more}
 P_{+-}=|\la m_+,z_-|\psi \ra|^{2}, 
\end{equation}
with similar expressions for the other three terms. A simple calculation shows that 
\begin{equation}\label{Ps}
P_{+-}=P_{-+}=\frac{1}{2}\cos^{2} \frac{\theta}{2}, \h
P_{++}=P_{--}=\frac{1}{2}\sin^{2} \frac{\theta}{2},
\end{equation}
 and from (\ref{infinitly correlation function}), we find:\\
\begin{equation}\label{correlation function}
Q_{\infty}=\cos\theta.
\end{equation}

If infinite singlet pairs were shared between Alice and Bob, then the value of $ Q_\infty $ doesn't show any fluctuation and Bob could find the exact value of the angle $\theta$ from (\ref{correlation function}).  Therefore while Alice is measuring along the ${\bf m}$ direction, Bob can make measurements along his $x$, $y$ and $z$ directions to determine the Euler angles of ${\bf m}$ and hence completely determine the vector ${\bf m}$ in the form  
\be \label{inft-estA}
{\bf m}=q_x \ {\bf x}+q_y \  {\bf y}+ q_z \ {\bf z}.
\ee
 If he has some  prior knowlege about the vector ${\bf m}$  being either in the upper or lower hemisphere, then he can determine the vector ${\bf m}$ by only two sets of measurements in the form 
\be \label{inft-estB}
{\bf m}=q_x \ {\bf x}+q_y \  {\bf y} \pm \sqrt{1-q_x^2-q_y^2} \ {\bf z},
\ee 
 where the sign is determined by the aforementioned prior knowledge. In the realistic case where the number of singlets is finite, then the above two methods, which we label as methods A and B respectively differ in their fidelity versus resource (i.e. number of singlets) used. We will make a detailed comparison of the two methods in the sequel. 

\subsection{The case where a finite number of pairs are shared}

In the realistic case where we have only a finite number $N$ of singlets, the correlations will fluctuate around their mean values and we can only estimate the vector ${\bf m}$. As the number of pairs increases the fluctuations decay and the fidelity of our estimation also increases.  To estimate the angle between two directions used by  Alice and Bob from correlations of their quantum measurements, we use the standard estimation procedure based on Bayesian inference \cite{Bayesian}. However, as in the ideal case, the rest of problem has a geometrical character and there are various methods for estimation of the final vector ${\bf m}$ and the final fidelity depends on our method of estimation. In any method used for estimation, the fidelity of the estimation between the original vector of Alice (${\bf m}$) and the estimated vector ${\bf m}_e$ is given by 
\be
F({\bf m}_e,{\bf m})=\frac{(1+{\bf m}\cdot {\bf m}_e)}{2}.
\ee
The average fidelity of this procedure is then given by
\be \label{avgFid}
\overline{F_N}:=\int d{\bf m}\int d{\bf m}_e P_N({\bf m}_e|{\bf m}) \frac{(1+{\bf m}\cdot {\bf m}_e)}{2},
\ee
where $P_N({\bf m}_e|{\bf m})$ is the conditional probability that the vector of Alice is ${\bf m}$ and it is estimated to be ${\bf m}_e$. Here $d{\bf m}=\frac{1}{4\pi}d\cos\theta \ d\phi$
with a similar expression for $d{\bf m}_e$.\\

We first consider estimation of the angle between one vector and ${\bf z}$ direction, say $\theta$ from measurement of correlations. Alice and Bob share $N$ singlets where Bob is measuring his qubits in his  ${\bf z}$ direction and Alice is measuring her qubits in a direction which appears as ${\bf m}$ in the coordinate system of Bob.  The correlation in this case is a random variable $Q_N$ which takes values $q_N$. In view of the relation (\ref{Correlation Function def}) and (\ref{Ps}), we have the conditional probability for the correlation to be $q_N$:
\be\label{ptheta}
P(q_N|{\bf z},{\bf m})=\binom{N}{{N_d}}(\cos^2\frac{\theta}{2})^{N_d}(\sin^2\frac{\theta}{2})^{N-{N_d}}.
\ee
Note that this probability depends only on the angle $\theta$. So it can equally be written as $P(q_N|\theta)$. Moreover it is easily seen from this binomial distribution that 
\be\label{mylabel}
\la q_N\ra=\cos \theta, \h \la q_N^2\ra=\cos^2\theta + \frac{1}{N}\sin^2\theta,
\ee
facts which will be used later on.  \\

{\bf Remarks on notation:} \\

{\bf i-} Since we take the measurement axis of Bob to be fixed along the ${\bf z}$ direction, we sometimes omit ${\bf z}$ from the conditional probabilities when there is no risk of confusion. \\

{\bf ii-} As it is evident from (\ref{Correlation Function def}), the correlation $q_N$ has a one-to-one correspondence with $N_d$, e.g. the number of times where Alice and Bob obtain opposite results in their measurements. In the following summations we use these two instead of each other, i.e. summing over $q$ from $-1$ to $1$ is equivalent to summing over $N_d$ from $0$ to $N$. \\

The conditional probability that  Alice has measured her spins along ${\bf m}$ given a specific value of correlation $q_N$ is given by   
\be
P({\bf m}\mid q_N)=\frac{P(q_N\mid {\bf m})P({\bf m})}{P(q_N)}=\frac{P(q_N\mid {\bf m})P({\bf m})}{\int d{\bf m} \ P(q_N\mid {\bf m})P({\bf m})}
\ee
where $P({\bf m})$ is the probability that Alice has measured her spins in the direction ${\bf m}$ and in the absence of any preference, this probability is taken to be uniform. 
From (\ref{ptheta}) we obtain 
\be
\int d{\bf m} P(q_N|{\bf m})=\frac{1}{4\pi}\int d\phi d\cos\theta \binom{N}{{N_d}}(\cos^2\frac{\theta}{2})^{N_d}(\sin^2\frac{\theta}{2})^{N-{N_d}}=\frac{1}{N+1},
\ee
where we have used the formula for Beta function
\be \label{beta function}
B(x+1,y+1)=\frac{x!y!}{(x+y+1)!}=\int d\theta (\cos\frac{\theta}{2})^{2x+1}(\sin\frac{\theta}{2})^{2y+1}.
\ee
This leads to  
\be \label{conditional probability}
P({\bf m}|q_N)=\frac{(N+1)!}{N_d!(N-N_d)!}(\cos^2\frac{\theta}{2})^{N_d}(\sin^2\frac{\theta}{2})^{N-{N_d}}.
\ee 

We now follow the standard estimation strategy and find the best estimate for ${\bf m}$ as:
\be
{\bf m}_{e}:=\int {\bf m} P({\bf m}\mid q_N) d{\bf m}.
\ee
or equivalently by using ${\bf z}\cdot {\bf m}=\cos \theta$ and ${\bf z}\cdot {\bf m}_{e}=\cos \theta_{e}$ 
\be
\cos\theta_{e}:=\int \cos\theta P({\bf m}\mid q_N) d{\bf m}.
\ee
Using (\ref{conditional probability}), a straightforward calculation  now gives 
\be\label{angle estimation}
\cos\theta_{e}=\frac{N}{N+2}q_N,
\ee
which goes to equation (\ref{correlation function}) for infinite $N$. 
This gives the estimated angle $\theta_e$ between ${\bf z}$ and ${\bf m}$ . \\

\section{Alignment of coordinate systems} \label{sec3}
The complete alignment of two coordinate systems is equivalent to the determination of the complete orientation of two orthogonal  vectors of Alice in Bob coordinate system. Therefore this problem reduces to the determination of one single vector of Alice  in Bob coordinate system. This part is purely geometrical to which we now turn.  To this end, we first study the case of two dimensions where Alice and Bob agree on a third direction (or a plane) and then go on to the full dimensional problem. \\

\subsection{Two dimensional coordinate systems}
\begin{figure}
\centering
\includegraphics[width=8cm,height=6cm]{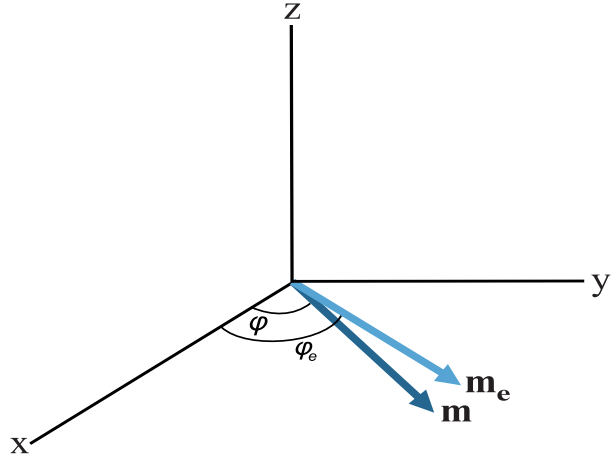}
\caption{Two dimensional coordinate sharing scheme, shown in the coordinate system of Bob: Alice and Bob agree on the $ {\bf z} $ direction and they want to share the direction $ {\bf m} $ which has been located in $ x-y $ plane. Bob estimates the desired direction $ {\bf m}_e $ due to the value of the correlation function.}
\label{2Dproblemfig}
\end{figure}

Here we assume that Alice and Bob agree on a third direction, say ${\bf z}$ and their problem is to align two $x-y$ coordinate systems in plane perpendicular to this direction, see figure (\ref{2Dproblemfig}).\\ 

The fidelity in this case is given by 
$F({\bf m}_e,{\bf m})=\frac{1+\cos (\phi-\phi_e)}{2}$ 
where $\phi$ and $\phi_e$ are respectively the actual and estimated angles of Alice vector with Bob $x$ axis. Using (\ref{avgFid}) we find the average fidelity to be 
\be
\overline{F}_N=\frac{1}{\pi}\int_{0}^{\pi}d\phi \sum_{N_d=0}^N
\binom{N}{N_d}
(\cos^2\frac{\phi}{2})^{N_d}(\sin^2\frac{\phi}{2})^{N-N_d}\frac{1+\cos(\phi-\phi_e)}{2} ,
\end{equation}
straight forward calculations by using Beta function (\ref{beta function}) give us: 
\be
\overline{F}_N= \dfrac{1}{2}+\dfrac{N}{4(N+2)}+\dfrac{1}{\pi(N+1)}\sum_{N_d=0}^N \sqrt{1-(\dfrac{2N_d-N}{N+2})^2}.
\ee
For large $N$, this leads to  
\begin{figure}
\centering
\includegraphics[width=12cm,height=6cm,angle=0]{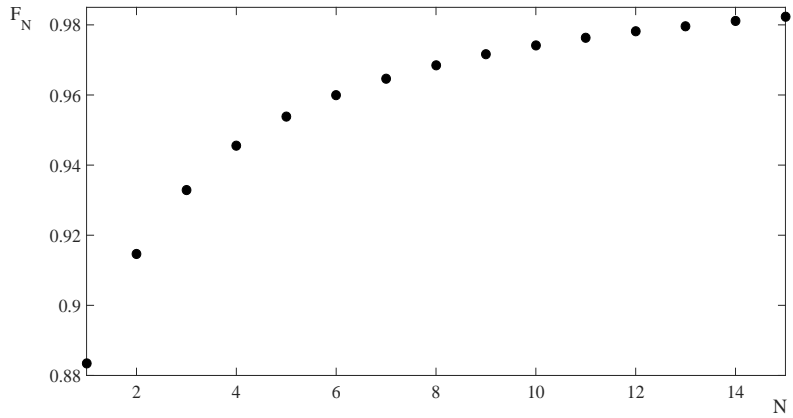}
\caption{The average fidelity for different values of $N$, when the direction $ {\bf m} $ which is estimated by Bob is located in a specific plane that he is aware of it.}
\label{fidelity2Dfig}
\end{figure}

\be
\lim_{N\rightarrow\infty} \overline{F}_N=\frac{3}{4}+\frac{1}{\pi}\int_{q=-1}^{1}\sqrt{(\dfrac{1-q}{2})(\dfrac{1+q}{2})} dq
=1 .
\ee
Figure (\ref{fidelity2Dfig}) shows the behavior of fidelity for different values of shared singlet pairs $ N $. As it can be seen in this figure, Alice and Bob can achieve the average fidelity around $0.9$ by using only three singlet pairs.\\

 \subsection{Three dimensional coordinate systems}\label{3DmethodA}
 We now come to the problem of aligning a full three dimensional coordinate system. This reduces to the problem of sharing three aligned directions. 
Similar to the procedure expressed in the ideal case of subsection (\ref{infinite case}), Alice and Bob may use $3N$ shared singlet pairs in method A, equation (\ref{inft-estA}), or use $2N$ singlet pairs as in method B, equation (\ref{inft-estB}). We explain in detail both methods and compare them with each other and also with the method of Massar and Popescu \cite{Massar and Popescu}.\\ 

\subsubsection{Method A; using $3N$ singlets}
\begin{figure}
\centering
\includegraphics[width=10cm,height=8cm,angle=0]{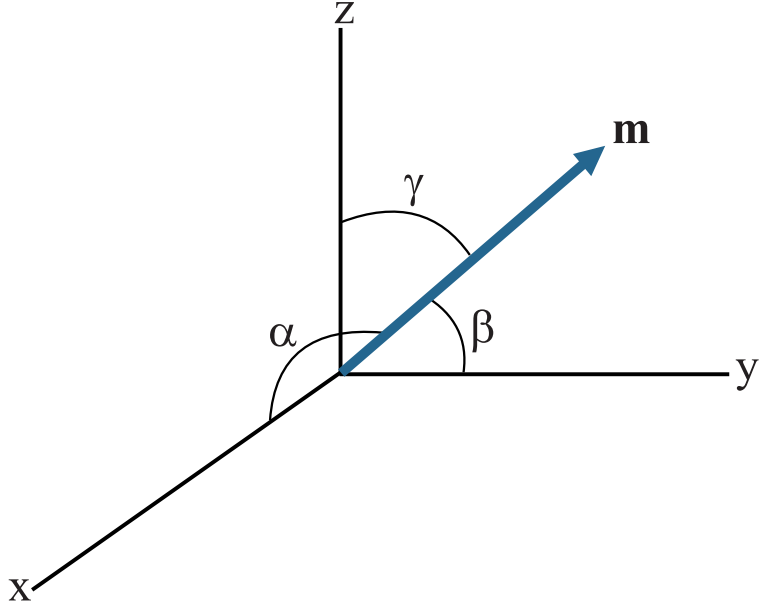}
\caption{Alice and Bob want to share the direction ${\bf m}$, which in the coordinate system of Bob it makes the angles $ \alpha $ , $ \beta $ and $ \gamma $ with $x$, $y$ and $z$ axes respectively.}
 \label{MethodA}
\end{figure}

Let $3N$ singlet pairs be shared between Alice and Bob and they want to share the original vector ${\bf m}$, which in the coordinate system of Bob has the form
\be
{\bf m}=\cos\alpha \ {\bf x} + \cos\b\  {\bf y}+\cos \gamma \ {\bf z},
\ee
as it is shown in figure (\ref{MethodA}). 
While Alice measures all her qubits along the ${\bf m}$ direction, Bob measures his qubits  in the directions ${\bf x}$, ${\bf y}$ and ${\bf z}$ ($N$ qubits in each direction) and finds the respective correlations $q_{x,_{N}}$, $q_{y,_{N}}$ and $q_{z,_{N}}$. 
Then by using equations (\ref{inft-estA}) and  (\ref{angle estimation}), he estimates ${\bf m}$ to be 

\be\label{mEstA}
{\bf m}_e = \frac{1}{\sqrt{q_{x}^2+q_{y}^2+q_{z}^2}}
\left(q_{x} {\bf x}
+q_{y} {\bf y}+q_{z} {\bf z}\right)
\ee
Note that in the above equation and hereafter we have dropped the subscript $N$ from $q$ for brevity, that is $q_x$ stands for $q_{x,_{N}} $.
Note that  due to fluctuations,  $q_{_{i}}$'s  are no  longer equal to the cosine of the angles of  ${\bf m}$ with the three axes, and hence the sum of their squares do not add to unity. Therefore normalization of the final vector is part of the estimation procedure in equation (\ref{mEstA}). \\

\begin{figure}
\centering
\includegraphics[width=13cm,height=6cm,angle=0]{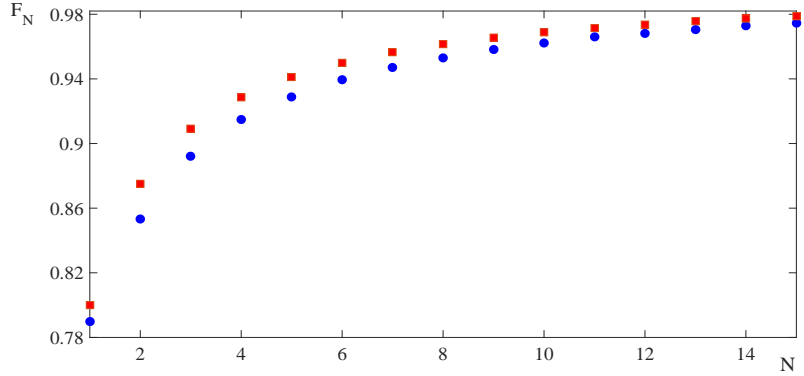}
\caption{(Color online) The blue circles show the behavior of the average fidelity for different values of N when 3N singlet pairs are shared between Alice and Bob and they use method A of section (\ref{3DmethodA}) to share a direction. The upper red squares show the optimal fidelity {\small{$\dfrac{3N+1}{3N+2}$}} which is achievable only by collective measurements \cite{Massar and Popescu}. We use $3N$ in the formula of optimal fidelity in order to have a comprehensive comparison.} 
 \label{fidelityMethodAfig}
\end{figure}

The probability of obtaining these correlations depends on the angles $\a$, $\beta$ and $\gamma$ as given by (\ref{ptheta}), see the paragraph after (\ref{ptheta}). Therefore we have

\be\label{probability of methodA}
P({\bf m}_e|{\bf m})=P(q_{_{x}},q_{_{y}},q_{_{z}}|{\bf m})=P(q_{_{x}}|\a)P(q_{_{y}}|\b)P(q_{_{z}}|\gamma),
\ee
where $P(q_{_{x,N}}|\a)$ is given by (\ref{ptheta}), with $\theta$ replaced by $\alpha$ and similar formulas for the other two probabilities hold. The fidelity between the vector and its estimate is given by $F({\bf m}_e,  {\bf m})=\frac{1}{2}(1+{\bf m}_e \cdot {\bf m})$ and the average fidelity,  is then given by
\be\label{average fidelity of methodA}
F^A_N= \sum_{q_x,q_y,q_z=-1}^1\int d{\bf m} P({\bf m}_e|{\bf m})F({\bf m}_e , {\bf m}).
\ee
%fidelity of methodA
Here the sum over $q_i$ from $-1$ to $1$ can be replaced with the sum over $N_{i,d}$ from $0$ to $N$, (see the second remark after equation (\ref{ptheta})) and the probability is given by (\ref{probability of methodA}). The right hand side of (\ref{average fidelity of methodA}) can be computed numerically and  the behavior of the average fidelity for different values of  $N$ can be seen in figure (\ref{fidelityMethodAfig}). One can see that our protocol achieves high fidelities for small values of $N$, even though we have used just one qubit measurements. The optimal fidelity of direction sharing is calculated in \cite{Massar and Popescu} when Alice sends the state $ |m\rangle^{\otimes N} $ to Bob and they show that the optimal measurement procedure necessarily involves a POVM on the whole system and can not be achieved by performing measurements on the components of the system, however we reach the optimal fidelity by only a very small gap. Figure (\ref{fidelityMethodAfig}) compares the average direction sharing fidelities of our protocol with that of the optimal method \cite{Massar and Popescu}, for different values of $N$. For example when $N=2$, Alice and Bob use $6$ singlet pairs and one qubit measurements to share the direction $ {\bf m} $ with average fidelity $ 0.85 $, whilst the optimal fidelity {\small{$\dfrac{6+1}{6+2}= 0.875 $}} can be achieved exclusively by global measurements. Hence if the laboratory restrictions force us to have simple measurements, our method will be a very good procedure for direction sharing.\\

\subsubsection{Method B; using $2N$ singlets}

Let $2N$ singlet pairs be shared between Alice and Bob. By the same procedure as in the ideal case of infinite number of pairs (sec. (\ref{infinite case})), while Alice is measuring along the ${\bf m}$ direction, Bob measures his first $ N $ qubits along the ${\bf x}$ axis and the other qubits along the ${\bf y}$ axis, and from the respective correlations $q_{_{x}}$ and $q_{_{y}}$ and by using equation (\ref{angle estimation}) he can estimate the vector ${\bf m}_e $ to be\\

\be\label{mEstC}
{\bf m}_e= \dfrac{Nq_{x}}{N+2} \ {\bf x}+\dfrac{N q_{y}}{N+2}\ {\bf y}+ \sqrt{1-(\dfrac{N q_{x}}{N+2})^2-(\dfrac{N q_{y}}{N+2})^2} \ \ \ {\bf z} ,
\ee
where we have used the partial information that the vector ${\bf m}$ lies in the northern hemisphere and hence have chosen the plus sign for $m_z$. \\

\begin{figure}
\centering
\includegraphics[width=13cm,height=6cm,angle=0]{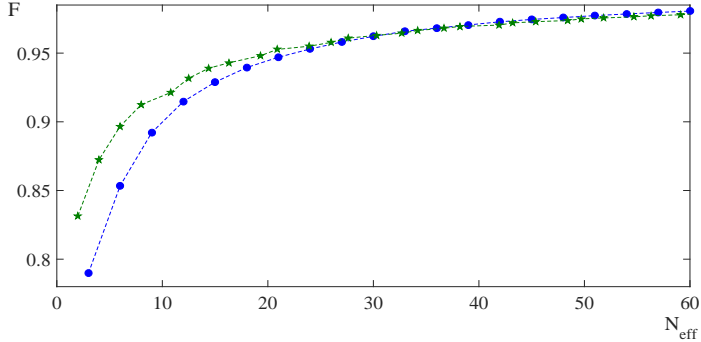}
\caption{(Color online) Average fidelity of estimation for two different methods as a function of the effective number of singlet pairs used in each run of the protocols. Blue circles correspond to the fidelities of method A and green stars correspond to the fidelities of method B. For more explanation see the paragraph after equation (\ref{goood}).}
 \label{comparsionfig}
\end{figure}

 Note that due to fluctuations of the values $q_{_{x}}$ and $q_{_{y}}$ for finite $N$, it may happen that $(\dfrac{Nq_{x}}{N+2})^2+(\dfrac{Nq_{y}}{N+2})^2 > 1$ in which case ${\bf m}_e$ cannot be defined. Such cases are inadmissible .  Bob has to abandon his inadmissible cases and repeat the protocol to obtain acceptable values for  $q_{_{x}}$ and $q_{_{y}}$ . The question is then what is the probability that Bob obtains inadmissible correlations, that is the probability of obtaining $(\dfrac{Nq_{x}}{N+2})^2+(\dfrac{Nq_{y}}{N+2})^2 > 1$. 
 To put a bound on this probability, we use Chebyshev formula according to which for a positive random variable $X$, we have
  \be
 Pr(X\geq a)\le \frac{\la X\ra}{a}.
 \ee
 In view of (\ref{mylabel}), this gives after some simple algebra
 
 \be
 Pr({inadmissible})\equiv Pr(q_x^2+q_y^2\geq (\frac{N+2}{N})^2)\leq (1-\cos^2\gamma + \frac{1}{N}(1+\cos^2\gamma))
 \ee
 where $\gamma$ is the angle between ${\bf m}$ and the $z-$ axis. 
 Obviously the probability depends on the angle $\gamma$. Averaging over all angles $\gamma$ in the northern hemisphere, this will give a bound
 
 \be\label{goood}
 \la Pr({inadmissible})\ra\leq (\frac{N}{N+2})^2\left(\frac{2}{3}+\frac{4}{3N}\right),
 \ee
 which shows that for large $N$ at least one-third of pairs lead to admissible correlations. In fact it is much better than this and numerical calculations show that the admissible probability is about $0.9$ for $N=15$. 
Since not all steps of the protocol of method B are admissible, in order to have a comprehensive comparision between methods A and B, we define the effective number of pairs used in each protocol to be $N_{eff}=\frac{N_{used}}{Pr(admissble)}$, for each method. $N_{used}$ is the number of pairs used in each run of the protocols.  In method A, $N_{used}=3N$ and $Pr(admissible)=1$, while for method B $N_{used}=2N$ and the admissible probability has been calculated numerically. Figure (\ref{comparsionfig}), shows the fidelities of both methods as a function of the effective number of pairs used. Note that although in method B,  Alice and Bob use $2N$ pairs, they rely on a prior information about their axes (for example being in the upper or lower hemisphere) and for small values of $N$ this prior information helps them to achieve higher fidelities compared with method A, see figure (\ref{comparsionfig}). As $N$ increases both methods reach the same fidelity which means that Alice and Bob compensate the lack of prior information in this case by consuming  extra shared singlets. For sufficiently large values of $N$, the fidelity of method A  exceeds that of method B, as expected.

  \section{Conclusion} \label{sec4}
In this article we have introduced a method for secure alignment of coordinate systems between two distant parties, in a way which prevents third parties from getting information about the aligned coordinate systems. This is certainly significant for the two parties who want to use measurements along different bases for performing quantum information tasks. The method is essentially based on obtaining information about directions from the correlations in the measurement results on singlet states shared between the two parties. The presented method achieves a very high average fidelity even though we have just one qubit measurements. In a sense this is the converse of what is done in Quantum Key Distribution, where instead of publicly announcing the measurement bases, Alice and Bob publicly announce the measurement results which enables them to align their coordinate systems. Of course they both have to trust the dealer who has send them truly singlet pairs. \\ 

The protocol is secure in the following sense: Clearly the access of Eve to the  classical strings $\left(a_1, a_2, \cdots , a_N \right)$ OR $\left(b_1, b_2, \cdots , b_N \right)$ publicly announced by Alice OR Bob, does not convey to her any information about the actual measurement directions of them. However the whole protocol can be sabotaged by Eve in the following ways. She can entangle herself with the entangled pairs shared by Alice and Bob, i.e. sharing a GHZ state with them in which case the correlations between Alice and Bob will be diminished considerably and will not lead to aligned reference frames. This kind of sabotage can be detected later by running  a test quantum information protocol by Alice and Bob. \\

 The other way that Eve can interfere is in the initial process of distributing singlets between the two parties.  For each singlet which is to be distributed and shared by Alice and Bob, Eve can intercept the qubit of say Alice, and instead produce a new singlet one qubit of which is kept by herself and the other one sent to Alice.   In this way she can share a singlet with Alice and another singlet with Bob. If in our protocol both Alice and Bob were to announce their classical strings of bits, then this would enable Eve to align two reference frames one with Alice and the other with Bob and preventing Alice and Bob to share an aligned reference frame. By further intercepting the classical communications between Alice and Bob, Eve could  interfere with any quantum information protocol  being run between Alice and Bob. However in our protocol only one of the parties announces his or her classical bits (results of measurements) while the other party keeps her or his results for comparison and determines the correlations. Therefore in this kind of attack, Eve can only align her reference frame with one of the parties. Further interception of classical messages between Alice and Bob does not help her anymore in hiding her presence which can  be detected  
by them once they perform a test quantum information protocol. \\

\section{Acknowledgements} 
 We would like to thank members of the QI group in Sharif University of Technology for their various useful comments on this work.

{}

\end{document}